\documentclass[aps,pra,reprint,superscriptaddress]{revtex4-1}

\usepackage{graphicx}
\usepackage{amsmath}
\usepackage{amsthm}
\usepackage{xcolor}

\def\AHU{School of Physics and Optoelectronic Engineering, Anhui University, Hefei, Anhui 230601, China}
\def\PKS{Max-Planck-Institut f\"{u}r Physik komplexer Systeme, D-01187 Dresden, Germany}
\def\ARU{Department of Physics and Astronomy, Aarhus University, DK-8000 Aarhus C, Denmark}
\begin{document}
	\title{Estimation of optimal control for two-level and three-level quantum systems with bounded amplitude}

	\author{Xikun Li}
	\affiliation{\AHU}
	\affiliation{\PKS}
	\affiliation{\ARU}

	\begin{abstract}
		A systematic scheme is proposed to numerically estimate the quantum speed limit and temporal shape of optimal control in two-level and three-level quantum systems with bounded amplitude. For the two-level system, two quantum state transitions are studied as illustration. Comparisons between numerical and analytical results are made, and deviations are significantly small. For the three-level system, two critical time points are determined with high accuracy, and optimal controls are obtained for different durations. The shape of optimized control field is simple and does not switch frequently, thus are easy to implement in experiment. In addition, we compare our method with the chopped random basis (CRAB), and the performance of our method is much better than that of CRAB. Our scheme is of importance in estimating the quantum speed limit and optimal control for cases in which analytical solution is absent. 
	\end{abstract}

	\maketitle
	
	
	\section{Introduction}
	
	Quantum optimal control (QOC) is crucial to engineer and to manipulate complex quantum systems in quantum information processing tasks~\cite{Glaser2015,Alessandro2021,Krotov1993,Brif2010}. Operations in experiments are generally executed very slowly, i.e., adiabatically, to avoid heating the sample and to guarantee the transition to the target state with perfect fidelity~\cite{Gericke2007}. However, the decoherence and noise in experiments sometimes make such slow operations impossible, thus it is desirable to achieve a speedup in a fast and robust manner~\cite{Odelin2019,Chen2010}. This is one important topic of quantum optimal control theory.
	
	Quantum optimal control theory has been widely applied in various physical systems such as Bose-Einstein condensate~\cite{vanFrank2016}, NMR~\cite{KHANEJA2005}, cold atoms in optical lattices~\cite{Li2018,Srivatsa2021}. One of the core problems in QOC theory is to find the time-optimal control. Time-optimal control problems focus on driving transitions to the target states in the \textit{minimal} time which is generally called the quantum speed limit (QSL)~\cite{Caneva2009}, and on finding the temporal shape of the control field. However, analytic solutions are only available for low-dimensional quantum systems~\cite{Lloyd2014}. For example, the analytic solutions are obtained for two-level system for energy minimization~\cite{Alessandro2001,Boscain2002} and time-optimal problems ~\cite{Khaneja2001,Boscain2005,Boscain2006,Boscain2014,Hegerfeldt2013,Hegerfeldt2014,Boozer2012,Jafarizadeh2020}. In addition, the minimal time and minimal energy problems in a three-level quantum system with \textit{complex} controls are studied with analytical solutions in Ref.~\cite{Boscain2002, Boscain2005}.
	
	For multiple-level quantum systems, in general, quantum optimal control relies on numerical methods, which employ local optimization algorithms, like Krotov~\cite{Sklarz2002}, GRAPE~\cite{KHANEJA2005}, CRAB~\cite{Doria2011}, GROUP~\cite{Sorensen2018} and GOAT~\cite{Machnes2018}, as well as global ones such as differential evolution (DE) and covariance matrix adaptation evolution strategy (CMA-ES)~\cite{Li2018,Zahedinejad2014}, and reinforcement learning~\cite{Bukov2018a}. It is worth noting that in experiments there might be constraints on the control field, e.g., the amplitude is bounded. In addition, the total duration is also limited for the time-optimal control problems. In such cases the local suboptimal traps exist in the quantum control landscape~\cite{Pechen2011,Larocca2018}. These local traps in general make the numerical methods fail to find the global optimal solution, even with the global optimization algorithms. Moreover, some methods cannot be applied in the problems where amplitude of control fields are bounded, while optimized controls found by some methods switch too frequently, or the temporal shape of control fields is too complicated, such it is hardly possible to implement them in real experiment.
	
	In this paper we propose a systematic scheme to numerically estimate the QSL and time-optimal controls. First, as an illustration, we investigate the time-optimal control problem for quantum state transitions in a two-level quantum system with bounded amplitude. Analytical results are provided in Refs.~\cite{Boscain2006}. We use the numerical scheme proposed to estimate QSL and time-optimal controls. In addition, the comparison between the numerical results and the analytical results is made. We find that deviations from the analytical results are significantly small. Furthermore, we compare our results with the results obtained with one state-of-the-art method, CRAB. The performance of our method is much better than that of CRAB. 
	
	Second, we employ this scheme in a three-level quantum system with \textit{real} control field which describes an external field \textit{without} rotating wave approximation (RWA). This problem is more difficult than the problems with complex controls studied in Ref.~\cite{Boscain2002, Boscain2005}. We estimate the optimal control for different values of total duration. Two regions of total durations are found in which the optimal control fields are of different types. QSL and time-optimal control are estimated. These results are compared with the ones obtained with CRAB.
	
	This paper is organized as follows. We define the control problem and propose a systematic scheme to estimate QSL in Sec.~\ref{sec:Model}. In Sec.~\ref{sec:twolevel} we consider two processes of quantum state transition in a two-level quantum system, and demonstrate the quantum control landscapes for certain types of control field. QSL and time-optimal control fields are estimated numerically, and these results are compared with the analytical results.  In Sec.~\ref{sec:threelevel} we study the optimal control problem of quantum state transition in a three-level quantum system. We conclude in Sec.~\ref{sec:conclusions}.	
	\section{Model}\label{sec:Model}
	
	\subsection{Hamiltonian and objective function}
	Here we consider the Hamiltonian with the control field $u(t)$:
	
	\begin{equation}
		H(t)  = H_0 + u(t) H_1
		\label{eq:bilinearH}
	\end{equation}	
	where $H_0$ is the drift Hamiltonian and $H_1$ is the control Hamiltonian (For simplicity, we consider the case in which there is only one control Hamiltonian). The bounded control field $u(t)$ is a real function under constraint $|u(t)|\leq M$. The dynamics of the system is governed by the controlled Hamiltonian $\mathrm{d} |\psi (t) \rangle / \mathrm{d}t = -\mathrm{i} H(t) |\psi (t) \rangle$, where we set $\hbar=1$.

	Numerically, for a quantum optimal control problem, we wish to optimize certain objective function such that the quantum speed limit $T_{\mathrm{qsl}}$ and optimized control field are obtained. The objective function is a functional of control field $u(t)$, thus we are able to investigate the control landscape for certain types of control field. One common choice of objective function is the fidelity between two quantum states. In our paper the fidelity is defined as follows:
	
	\begin{eqnarray}\label{eq:Fidelity}
		F(u(t),T) & =|\langle\psi_{t}|\mathcal{T}\exp (- \mathrm{i}\int_0^T H(t) \mathrm{d}t
		|\psi_{i}\rangle|^2 \\ \nonumber 
		=|\langle\psi_{t}|\psi_{f}\rangle|^2.
	\end{eqnarray}
	where $\mathcal{T}$ is the time-ordering operator, and $|\psi_{t}\rangle$ ($|\psi_{i}\rangle$) is the target (initial) state. $T$ is the total duration of time evolution, and $|\psi_{f}\rangle$ is the final state after the time evolution. An alternative characterising the similarity between the final state $|\psi_{f}\rangle$ and the target state$|\psi_{t}\rangle$ is the Bures distance
	\begin{equation}
		d_B=\sqrt{2 (1-|\langle\psi_{f} |\psi_{t}\rangle|}=\sqrt{2 (1-\sqrt{F})}.
		\label{eq:Bures}
	\end{equation}
	Perfect fidelity gives zero Bures distance, and vice versa.

	\subsection{Bang-Off control}
	
	Inspired by Ref.~\cite{Boscain2006} we consider the bang-off control as the control protocol to be optimized. For certain duration $[s,s+\delta s]$, if the control is restricted to take its maximum (minimum) $u(t)=M$ ($u(t)=-M$), it is called a \textit{bang} control, and is denoted by $P_{\delta s}$ ($N_{\delta s}$), here $P$ ($N$) is short for Positive (Negative); Similarly, if the value of control is zero $u(t)=0$, it is called a \textit{off} control, and is denoted by $0_{\delta s}$. Note that in the case of two-level system the off control is \textit{singular} control~\cite{Boscain2006}. The bang-off control refers to a finite concatenation of bang and off controls.   
	
	We consider two classes of bang-off control. For the first class the duration for each bang (and off) control is in general different, e.g., $P_{t_1} 0_{t_2} N_{t_3}$ defined in the following
	
	\begin{equation}
		u(t)= \Bigg\{ \begin{matrix}
			M& 0 \;\;\leq t < t_1 \\
			0 & \;\; t_1\leq t < t_1+t_2 \\
			-M & \;\; t_1+t_2\leq t \leq t_1+t_2+t_3, \\
		\end{matrix}
		\label{eq:bang1}
	\end{equation}
	where the order of letter sequence is from left to right. For the example above, the bang-off control is switched from bang ($P$) to off ($0$), then to bang ($N$), with a switch number being two $N_s=2$. The \textit{type} of such bang-off controls refers to a way in which bang and off controls are concatenated. For a given number of switches $N_s$, the number of possible types $N_{type}$ is at most $3\times 2^{N_s}$. For certain initial/target quantum states, $N_{type}$ can be further reduced. The control fields are represented by the type and vector of durations $\mathbf{t}=[t_1,t_2,...]$. 
	
	For the second class the duration of each bang/off is the time step-size: $\delta t=T/N_T$ with $T$ being total duration and $N_T$ the number of time slots. Thus the control is usually represented by a control vector $u(t)\equiv \{u_k| k=1,...,N_T\}$. This class of bang-off control arises because of a coarse-graining of temporal variable for a numerical optimization. In general, the simulation is more accurate when $N_T$ is larger. However, the optimization with large $N_T$ is time-consuming, and the dimension of search space is huge. For the problems considered here the control is allowed to take 3 different values $u_k=\{\pm M,0\}$, thus the possible number of control vectors is $3^{N_T}$, which is an exponentially large number. For the quantum optimal control problem of two-level system, $N_s$ is much smaller than $N_T$ in general. Thus the optimization of first class of bang-off controls is much easier than that of second one. In addition, the optimized controls of second class generally switch more frequently than the ones of first class, thus are less desirable in real experiments. 
	
	Note that we do not claim that the first class of bang-off control is time-optimal control protocol for \textit{all} QOC problems, e.g., time-optimal control of nonlinear two-level quantum system~\cite{Chen2010}. Instead, we use the first class of bang-off control to \textit{estimate} QSL and to \textit{approximate} the temporal shape of time-optimal control. This is due to the fact that the shape of first class of bang-off control is simple and its universality is in general better than bang-bang control. In addition, the optimality of first class of bang-off control has been proved in the two-level Landau-Zener system~\cite{Boscain2006}. Therefore, we mainly focus on the study of first class of bang-off control.

	\subsection{Optimization algorithm}
	
	Optimization algorithms are of great importance for quantum optimal control problems. Two common choices are gradient-based and gradient-free algorithms. For the gradient-based algorithm, we choose the quasi-Newton method. For the gradient-free algorithm, we use stochastic descent (SD) as the optimization algorithm~\cite{Bukov2018a}. Stochastic descent starts from a random configuration and updates control field $u(t)$ only if the candidate increases (decreases) the fidelity (Bures distance).

	\subsection{Scheme}
	Specifically, for the first class of bang-off controls, we start from $N_s=0$, and look through all possible types. For each type we optimize using SD (or quasi-Newton) the vector of durations $\mathbf{t}$ for different values of total duration $T$, until the perfect fidelity $F=1-\delta$ is found, where $\delta$ is vanishingly small. We denote $T_i^{\mathrm{min}}$ the minimal duration for which the perfect fidelity $F=1- \delta$ is obtained with number of switches being $N_s=i$. If $T_i^{\mathrm{min}} \approx T_{i+1}^{\mathrm{min}}$ and it is zero that one duration of optimized duration vector $\mathbf{t}$ with $N_s=i+1$, then this means the increment of number of switches does not decrease the minimal duration to reach perfect fidelity. Hence we can stop optimization and choose the types corresponding perfect fidelity with optimized $\mathbf{t}$ and $N_s=i$ as the estimation of time-optimal control, and $T_i^{\mathrm{min}}$ as the estimation of QSL.
	
	For the second class, we optimize the control vectors using 1-flip SD. Here 1-flip means that the new candidates are generated by randomly choosing \textit{one} time-slot and changing the value of control at that time-slot to other values~\cite{Bukov2018a}. 
	
	\section{Two-level system}\label{sec:twolevel}
	
	Here we consider a two-level quantum system driven by the Hamiltonian described in the following:
	
	\begin{equation}
		H(t)  =-E \sigma_z + u(t) \sigma_x 
		\label{eq:Hamiltonian}
	\end{equation}
	where $\sigma_z$ is the drift Hamiltonian and $E>0$, and $\sigma_x$ is the controlled Hamiltonian. When the control field is turned off, the lower (upper) eigenstate is $|0\rangle= [1, 0]^{\mathrm{T}}$ ($|1\rangle= [0, 1]^{\mathrm{T}}$).
	
	One important parameter is defined as: $\alpha \equiv \arctan(M/E)$. The reason is in the following. For $\alpha<\pi/4$ the optimal control is bang-bang, and the durations for the first, final and interior bang have to be determined numerically; for $\alpha>\pi/4$ the optimal control is in general \textit{bang-off} control with at most two switches~\cite{Boscain2006}. For the latter case the bang-bang fails to capture time-optimal control, and this is the case we are interested. Therefore, we choose $E=1$ and $M=4/3$ such that $\alpha=0.9273>\pi/4$.
	
	We are interested in two transition processes with the bounded controls: (1) transition from $|0\rangle$ to $|1\rangle$; (2) transition from $|0\rangle$ to $|\psi_E\rangle$, where $|\psi_E\rangle= 1/\sqrt{2}(|0\rangle + e^{i 9\pi/10}|1\rangle)$. For the first case the time-optimal control is in fact bang-bang, whereas for the second the time-optimal control is bang-off type. We wish to estimate the quantum speed limit, as well as to determine the temporal shape of the time-optimal controls, of two transitions with the aid of numerical optimization over bang-off control fields.
	
	\subsection{Transition from $|0\rangle$ to $|1\rangle$}\label{sec:case1}
	
	\begin{figure}
		\includegraphics[width=1\linewidth, trim= 100 230 100 250,clip]{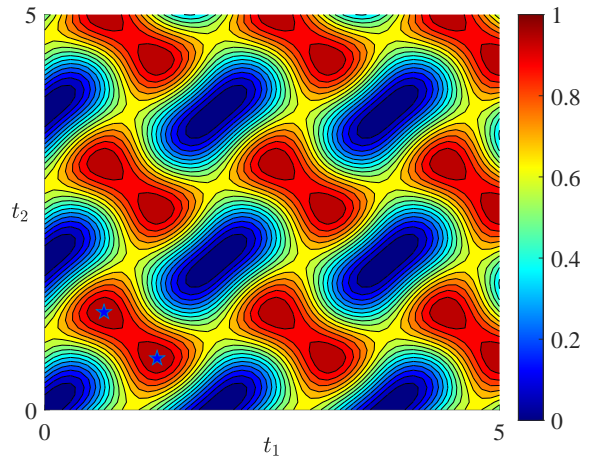}
		\caption{Quantum control landscape of fidelity $F$ for type $P_{t_1}N_{t_2}$ of the first class bang-off controls with initial state $|0\rangle$ and target state $|1\rangle$. Two time-optimal controls are indicated by two blue pentagrams with $[t_1,t_2]=\{[T_1,T_2],[T_2,T_1]\}$.
		}
		\label{fig:landscape1}
	\end{figure}
	
	In Ref.~\cite{Boscain2006} it has been proved, for initial state $|0\rangle$ and target state $|1\rangle$ with $\alpha>\pi/4$, the time-optimal controls are the first class of bang-off controls with one switch $N_s=1$, i.e., $\{P_{T_1}N_{T_2}, P_{T_2}N_{T_1}, N_{T_1}P_{T_2}, N_{T_2}P_{T_1}\}$, where $T_1= (\pi - \arccos{(\cot^2{\alpha})})/2 \sqrt{E^2+M^2}=0.6505$ and $T_2=(\pi + \arccos{(\cot^2{\alpha})})/2 \sqrt{E^2+M^2}=1.2345$. Thus the quantum speed limit is $T_{\mathrm{QSL}}=T_1+T_2=\pi/\sqrt{E^2+M^2}=0.6\pi$.
	
	We wish to numerically estimate $T_{\mathrm{QSL}}$ and to obtain the time-optimal control field using the scheme proposed. We first investigate the first class with $N_s=1$, and show that the minimal duration for perfect fidelity is equal to $T_{\mathrm{QSL}}$. Then we show the bang-off controls with $N_s=2$ reduce to the ones with $N_s=1$ for the time-optimal problem considered here. At last, we consider the second class of bang-off controls, and compare these results with the ones obtained using CRAB method. 
	
	\subsubsection{First class with $N_s=1$}
	
	\begin{figure}
		\includegraphics[width=1\linewidth, trim= 60 230 50 250,clip]{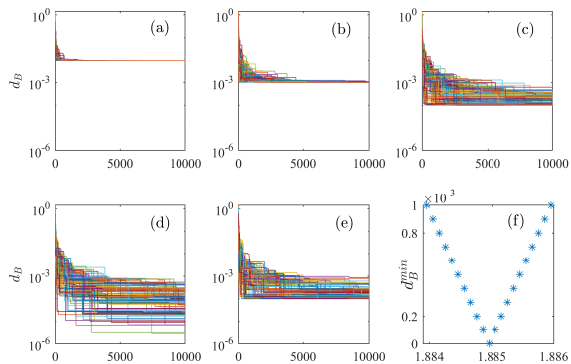}
		\caption{Bures distance $d_B$ versus iterations for different values of total evolution duration $T$ for type $P_{t_1}N_{t_2}$: (a) $T=T_{\mathrm{QSL}}-10^{-2}$; (b) $T=T_{\mathrm{QSL}}-10^{-3}$; (c) $T=T_{\mathrm{QSL}}-10^{-4}$; (d) $T=T_{\mathrm{QSL}}$; (e)  $T=T_{\mathrm{QSL}}+10^{-4}$. The type of bang-off control is $P_{t_1}N_{t_2}$ with initial state $|0\rangle$ and target state $|1\rangle$, and the optimization algorithm is SD. The number of initial search points is one hundred, and the number of iteration is ten thousand. (f) The minimal $d_B^{\mathrm{min}}$ obtained after 10000 iterations for different values of total duration $T$. When $T=T_{\mathrm{QSL}}$, the sequence of $d_B^{\mathrm{min}}$ reaches its minimum. 
		}
		\label{fig:iteration}
	\end{figure}
	
	We start from the first class of bang-off controls with $N_s=1$. For the transition from  $|0\rangle$ to $|1\rangle$, the number of possible types is two, i.e., $\{P_{t_1}N_{t_2}, N_{t_1}P_{t_2}\}$.
	
	In Fig.\ref{fig:landscape1} we show the quantum control landscape for type $P_{t_1}N_{t_2}$. The control fields are characterized by a two-dimensional vector of durations $\mathbf{t}=[t_1,t_2]$. A similar landscape is obtained for type $N_{t_1}P_{t_2}$. The repetition of landscape over the square domain with time length $T_{\mathrm{QSL}}$ is because the time evolution of quantum state is periodic. In Fig.1 no local traps exist, while it is not true for the second class of bang-off controls. This is one of the advantages of the first class over the second one.
	
	We now estimate the minimal duration $T_1^{\mathrm{min}}$ for which the perfect fidelity $F=1- \delta$ is obtained. The procedure is in the following: choose different values of $T$ with $t_1+t_2=T$; optimize the free variable $t_1$ using SD to maximize (minimize) fidelity $F$ (Bures distance $d_B$). When $T \neq  k T_{\mathrm{QSL}}, k=1,2,...$ we find that $d_B$ does not converge to zero. In addition, the optimization stops within fewer iterations when $T$ deviates more from $T_{\mathrm{QSL}}$. When $T=T_{\mathrm{QSL}}$, however, $d_B$ keeps decreasing and the update of control fields continues for a large number of iterations.

	In Fig.~\ref{fig:iteration}(a)-(e) we show the Bures distance as a function of iteration for type $P_{t_1}N_{t_2}$ with five different values of $T$. One hundred initial search points of $t_1$ are randomly generated, and are optimized using SD. We find that as $T$ approaches toward $T_{\mathrm{QSL}}$, the convergence rate decreases. Moreover, the minimal Bures distance obtained after optimization decreases as $T$ approaches toward $T_{\mathrm{QSL}}$, cf. Fig.~\ref{fig:iteration}(f). When $T=T_{\mathrm{QSL}}$, the Bures distance as a function hardly converge even for very large number of iterations. Therefore, we could estimate $T_{\mathrm{QSL}}$ by monitoring the convergence of Bures distance as an alternative method. For instance, in Fig.~\ref{fig:iteration}(c) $T=T_{\mathrm{QSL}}-10^{-4}$, the search point with the minimal $d_B$ found in the last iteration stops decreasing after several hundred iteration. For $T=T_{\mathrm{QSL}}$, however, the optimization continues after ten thousand iterations, as shown in Fig.~\ref{fig:iteration}(d). Therefore, the minimal duration with one switch is approximately equal to the quantum speed limit $T_1^{\mathrm{min}}=T_{\mathrm{QSL}}$ with very high precision. In Fig.~\ref{fig:iteration}(d) 100 optimized values of $t_1$ cluster around $T_1$ and $T_2$ with significantly small standard deviation $\sigma \sim \mathcal{O}(10^{-4})$. The best fidelity using quasi-Newton method is $F=1-\mathcal{O}(10^{-16})$. Same results are obtained for type $N_{t_1}P_{t_2}$. Therefore the numerical results reproduce the analytical solutions with very high accuracy. 
	
	\begin{figure}
		\includegraphics[width=1\linewidth, trim= 10 230 10 250,clip]{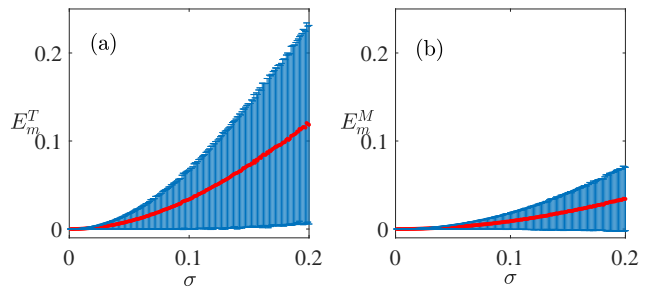}
		\caption{Error distributions of $E=1-F$ as a function of standard deviation $\sigma$ for one time-optimal control $P_{T_1}N_{T_2}$. Red dots are the mean, and the vertical blue lines represents the standard deviation of $E$. (a) Error $E_m^T$ resulting from the uncertainty of time interval for each bang control. 10000 random duration vectors $\mathbf{t}$ are drawn from Gaussian distribution with mean being $[T_1,T_2]$ and standard deviation $\sigma$. (b) Error $E_m^M$ resulting from the uncertainty of boundaries $\pm M$ for each bang control. 10000 random values of boundaries are drawn from Gaussian distribution with mean being $\pm M$ and standard deviation $\sigma$.
		}
		\label{fig:error}
	\end{figure}
	
	From the experimental point of view, the errors arise due to the operational inaccuracy. Therefore, we investigate the robustness of time-optimal bang-off controls with respect to uncertainty in the duration vector $\mathbf{t}$ and in the boundary of control field $\pm M$. We sample 10000 duration vectors $\mathbf{t}$ and calculate the corresponding error distribution $E_m^T$ for different values of $\sigma$. Similar is done for $E_m^M$. In Fig.~\ref{fig:error} we show $E_m^T$ and $E_m^T$ as a function of standard deviation $\sigma$. The mean of $E$ is roughly a constant multiply the square of $\sigma$. Same for standard deviation of $E$. In addition, by comparing two plots in Fig.~\ref{fig:error} we observe that the time-optimal control is less robust w.r.t. the uncertainty of duration vector than that of boundaries.

	\subsubsection{First class with $N_s=2$}
	\begin{figure}
		\includegraphics[width=1\linewidth, trim= 100 230 100 250,clip]{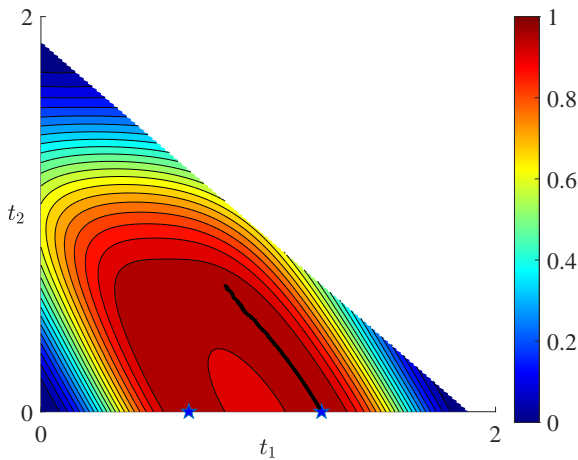}
		\caption{Quantum control landscape of fidelity $F$ for type $P_{t_1}0_{t_2}N_{t_3}$ with initial state $|0\rangle$ and target state $|1\rangle$, and $t_1+t_2+t_3=T_{\mathrm{QSL}}$. Two time-optimal controls are indicated by two pentagrams for which $[t_1,t_2]=\{[T_1,0],[T_2,0]\}$ The black dots indicate the optimization process of one search point which climbs towards one of optimal time vector $[T_2, 0]$ with the global maximum of fidelity $F=1$.
		}
		\label{fig:landscape2}
	\end{figure}
	
	To numerically verify that the time-optimal controls are indeed of type with $N_s=1$ for the case considered in this section, we continue to study the first class of bang-off controls with $N_s=2$. We would show that $T_2^{\mathrm{min}}=T_1^{\mathrm{min}}$ with high precision, and one of three durations is zero, thus controls with $N_s=2$ reduce to the ones with $N_s=1$.
	
	Specifically, for the transition from  $|0\rangle$ to $|1\rangle$, the number of possible types is six, i.e., $\{P_{t_1}0_{t_2}N_{t_3}, N_{t_1}0_{t_2}P_{t_3}, P_{t_1}N_{t_2}P_{t_3}, N_{t_1}P_{t_2}N_{t_3}, P_{t_1}0_{t_2}P_{t_3}, \newline N_{t_1}0_{t_2}N_{t_3}\}$. For the first four types, the perfect fidelity are reached when $T = T_{\mathrm{QSL}}$, and one of the durations is zero, e.g., $t_2=0$ for $P_{t_1}0_{t_2}N_{t_3}$. For the last two types, the maximal fidelity is $F=0.6399$ when $T = T_{\mathrm{QSL}}$. When $T<T_{\mathrm{QSL}}$, the perfect fidelity cannot be reached for all six types. For instance, the maximal fidelity with $T=T_{\mathrm{QSL}}-0.001$ is $F=1-1.001\times10^{-6}$ for type $P_{t_1}0_{t_2}N_{t_3}$. Similar results are obtained for other types. Therefore, we have numerically verified that $T_2^{\mathrm{min}}=T_1^{\mathrm{min}}=T_{\mathrm{QSL}}$.
	
	In Fig.~\ref{fig:landscape2} we show the quantum control landscape for type $P_{t_1}0_{t_2}N_{t_3}$ with $t_1+t_2+t_3=T_{\mathrm{QSL}}$. Two time-optimal controls are indicated by two pentagrams with $[t_1,t_2]=\{[T_1,0],[T_2,0]\}$. The duration for off-control $u(t)=0$ is zero $t_2=0$, thus controls with $N_s=2$ reduce to the ones with $N_s=1$. The optimization process of one search point using SD climbs towards one of the optimal time vector. As search point is climbing towards the global maximum, smaller step-size and more iterations are required for better performance.
	
	We optimize the time vector $\mathbf{t}=[t_1, t_2]$ for type $P_{t_1}0_{t_2}N_{t_3}$ with $t_3=T_{\mathrm{QSL}}-t_1-t_2$. The best fidelity obtained using quasi-Newton is $F=1-\mathcal{O}(10^{-16})$, and the optimized duration vectors deviate from the optimal values, e.g., $\mathbf{t}_{\mathrm{Opt}}=[T_2,0,T_1]=[1.2345, \,0, \,0.6505]$ with vanishingly small value. Similar results are obtained for types $\{N_{t_1}0_{t_2}P_{t_3}, P_{t_1}N_{t_2}P_{t_3}, N_{t_1}P_{t_2}N_{t_3}\}$.
	
	The increment of switch numbers $N_s=1 \rightarrow N_s=2$ does not result in the decrement of total duration, but $T_2^{\mathrm{min}}=T_1^{\mathrm{min}}=T_{\mathrm{QSL}}$. In addition, we have shown that one duration of the optimized time vector is approximately equal to zero, thus the controls with $N_s=2$ reduce to the ones with $N_s=1$. Then we can stop studying the controls with larger number of switches $N_s \geq 3$. 
	
	In summary we have numerically verified for the transition $|0\rangle  \rightarrow |1\rangle$ the time-optimal controls are $\{P_{T_1}N_{T_2}, P_{T_2}N_{T_1},N_{T_1}P_{T_2},N_{T_2}P_{T_1}\}$, and the quantum speed limit is $T_{\mathrm{QSL}}$, with little deviations.
	
	\subsubsection{Second class of bang-off control}
	
	\begin{figure}
		\includegraphics[width=1\linewidth, trim= 60 200 90 230,clip]{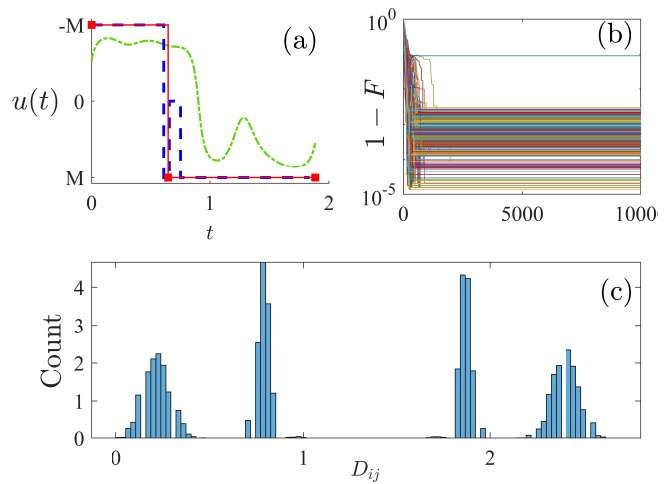}
		\caption{The results obtained for transition from $|0\rangle$ to $|1\rangle$ with $T=T_{\mathrm{QSL}}$ using 1-flip SD to optimize the second class of bang-off controls. The number of time slots is fixed to be $N_T=40$. (a) The optimized control field found with first class of bang-off (red,solid) which is $P_{T_1}N_{T_2}$, that with second class (blue, dashed), and with CRAB (grean, dash-dotted line). (b) Infidelity $1-F$ versus iterations with 1000 search points. (c) Distribution of distance $D_{ij}$ between optimized control fields. 
		}
		\label{fig:distance_1}
	\end{figure}
	
	In this sub-section we consider the second class of bang-off controls as a comparison. Here we employ the 1-flip SD to optimize the control vectors where each component is allowed to take one of three values $u_k=\{\pm M, 0\}$. The number of time slots is fixed to be $N_T=40$. 1000 initial control vectors are randomly generated and are optimized using SD for $10^4$ iterations. To further investigate the relations between the optimized control fields, we calculate the distance between optimized control fields as 
	
	\begin{equation}
		D_{ij}=\frac{1}{N_T}||u^{(i)}-u^{(j)}||_1
		\label{eq:distance}
	\end{equation}
	where $u^{(i)}$ is the $i$-th optimized control field and $||\cdot||_1$ is the absolute-value norm~\cite{Larocca2018}.  
	
	In addition, we compare the results to one of the state-of-the-art algorithms, Chopped Random Basis (CRAB). We choose various value of $N_c$ which is the cut-off number  of Fourier basis, and optimize the corresponding control field with $T=T_{\mathrm{QSL}}$. The best fidelity obtained is $F=1-\mathcal{O}(10^{-15})$ with $N_c=5$.
	The corresponding optimized control field is shown in Fig.~\ref{fig:distance_1}(a) by green dash-dotted line.
	
	In Fig.~\ref{fig:distance_1}(a) the optimized control field found with second class is indicated by the blue dashed line, while one of the time-optimal control fields $P_{T_1}N_{T_2}$ is marked by the red solid line. We observe that the temporal shape of the optimized control field found by 1-flip is similar to that of time-optimal control $P_{T_1}N_{T_2}$, but the number of switch is three. Both the optimized control fields obtained with second class of bang-off and that with CRAB approximate reasonably the time-optimal control field. 
	
	In Fig.~\ref{fig:distance_1}(b) we plot the infidelity $1-F$ as a function of optimization iteration. 1-flip SD stops updating after about 1000 iterations, and the maximal fidelity obtained is $F=1-1.35\times 10^{-5}$. Presumably, the performance of second class of bang-off controls would be improved by $k$-flip SD ($k \geq 2$), yet the number of iterations required is much larger~\cite{Bukov2018a}. In Fig.~\ref{fig:distance_1}(c) distances between optimized control fields $D_{ij}$ form a multimodal distribution with four peaks. This yields the fact that the optimized control fields cluster around four time-optimal controls $\{P_{T_1}N_{T_2},P_{T_2}N_{T_1},N_{T_1}P_{T_2},N_{T_2}P_{T_1} \}$ with a standard deviation much larger than that obtained with first class of bang-off controls. 
	
	Therefore, for the time-optimal problem of quantum transition from $|0\rangle$ to $|1\rangle$, the performance of first class of bang-off control is much better than that of second class optimized using 1-flip SD, concerning the maximal fidelity obtained, the optimized control fields found and the their distribution.

	\subsection{Transition from $|0\rangle$ to $|\psi_E\rangle$ }\label{sec:case2}

	In this section we consider the transition from $|0\rangle$ to $|\psi_E\rangle$ with $\alpha > \pi/4$. The time-optimal controls are of first class bang-off controls with two switches $N_s=2$: $\{P_{\tau_1} 0_{\tau_2} N_{\tau_3}, P_{\tau_1}0_{\tau_2}P_{\tau_3} \}$~\cite{Boscain2006}. Three durations are: $\tau_1=T_1=0.6505$ is the duration for which the time-evolving quantum state reaches the equator of Bloch sphere for the first time under the bang control $u(t)=M$. The quantum state after this process is $|A^+ \rangle =1/\sqrt{2}(|0\rangle + e^{i(\pi+\beta)}|1\rangle)$ (up to a global phase) with $\beta=\arccos{E/M}=0.7227$; $\tau_2=(\beta-\pi/10)/2E=0.2043$ is the duration for which the quantum state reaches $|\psi_E '\rangle= 1/\sqrt{2}(|0\rangle + e^{i 11\pi/10}|1\rangle)$ along the equator starting from $|A^+ \rangle$ with the off control $u(t)=0$; and $\tau_3=0.2978$ is the duration for which the quantum state reaches target state $|\psi_E\rangle$ from $|\psi_E '\rangle$ with the control fields $u(t)=\pm M$. The quantum speed limit is $\tau_\mathrm{QSL}=\tau_1+\tau_2+\tau_3=1.1525$. Here we use a different notation $\tau_\mathrm{QSL}$ for QSL to distinguish it from the one in Sec.~\ref{sec:case1}. 
	
	\subsubsection{First class with $N_s=1$}
	
	For the transition considered in this section, the possible types of first class are four: $\{P_{t_1}0_{t_2}, P_{t_1}N_{t_2},N_{t_1}0_{t_2}, N_{t_1}P_{t_2}\}$. When $T=\tau_\mathrm{QSL}$, the maximal fidelity of four types are: $F=[0.9997,1-5.0937\times10^{-6},0.4999,0.9873]$. Therefore, the time-optimal controls cannot be the ones with $N_s=1$ because $T_1^{\mathrm{min}}>\tau_\mathrm{QSL}$. It is worth noting that although $T_1^{\mathrm{min}}>\tau_\mathrm{QSL}$, the perfect fidelity can be obtained using $N_s=1$ controls with a larger duration, e.g., $P_{\tau_1}0_{\lambda}$ with $\lambda=\beta+\pi/10=0.5184$ and the total duration $T=\tau_1+\lambda=1.1689$ which is a little larger than $\tau_\mathrm{QSL}=1.1525$. Because of the operational inaccuracy in experiment, the larger $N_s$ is, the larger the error is. Therefore, the controls with $N_s=1$ is acceptable, provided the perfect fidelity can be obtained and the total duration which does not exceed $\tau_\mathrm{QSL}$ much is tolerable. 
	
	\subsubsection{First class with $N_s=2$}
	\begin{figure}
		\includegraphics[width=1\linewidth, trim= 100 230 100 250,clip]{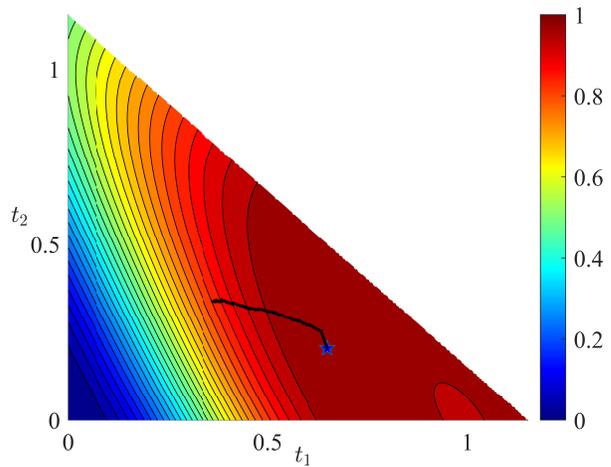}
		\caption{Quantum control landscape of fidelity $F$ for type $P_{t_1}0_{t_2}N_{t_3}$ with initial state $|0\rangle$ and target state $|\psi_E\rangle$, and $t_1+t_2+t_3=\tau_{\mathrm{QSL}}$. The time-optimal control is indicated by a blue pentagram with $\mathbf{\tau}_{\mathrm{Opt}}=[\tau_1,\tau_2]$. The black dots indicate the optimization process of one search point which climbs towards one of optimal time vector $\mathbf{t}_{\mathrm{Opt}}$ with the global maximum of fidelity $F=1$.
		}
		\label{fig:landscape3}
	\end{figure}
	
	\begin{figure}
		\includegraphics[width=1\linewidth, trim= 120 150 150 150,clip]{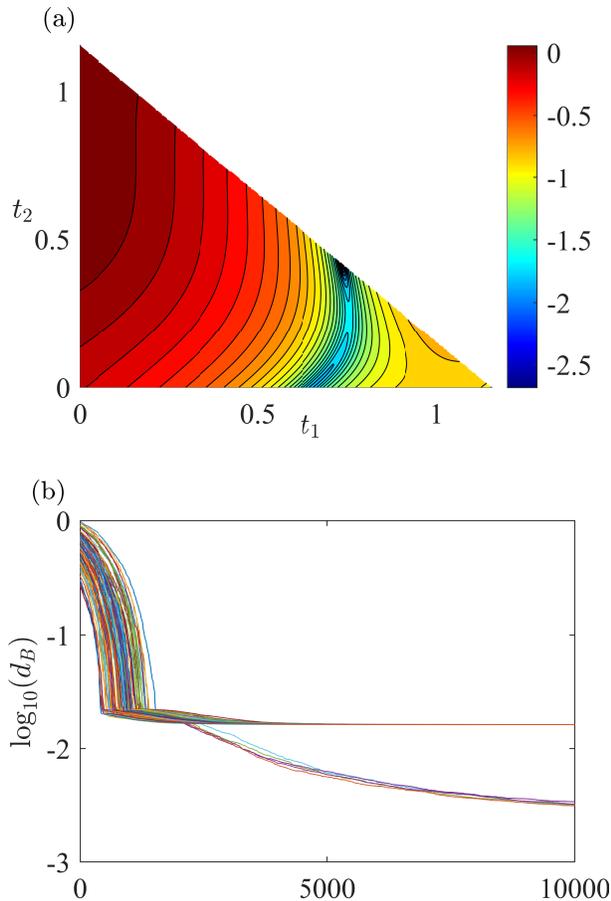}
		\caption{(a) Quantum control landscape of log-Bures distance $\log_{10}d_B$ for type $P_{t_1}N_{t_2}0_{t_3}$ with $t_1+t_2+t_3=\tau_{\mathrm{QSL}}$.Two local minima are shown with blue color. (b) $\log_{10}d_B$ as a function of optimization iteration with 100 initial search points. 100 search points saturated towards two branches with values of $\log_{10}d_B$ being $[-1.7916,-2.6825]$ after 10000 iterations.
		}
		\label{fig:PN0}
	\end{figure}
	
	We could estimate $T_2^{\mathrm{min}}$ using the same scenario in Sec.~\ref{sec:case1}. An alternative method is to calculate the fidelity for time vectors in the fine-grained $3D$ temporal space. Find the locations of perfect fidelity and choose the shortest duration as the estimate of $T_2^{\mathrm{min}}$.

	In Fig.~\ref{fig:landscape3} we show the quantum control landscape of fidelity for $P_{t_1}0_{t_2}N_{t_3}$ with $t_1+t_2+t_3=\tau_{\mathrm{QSL}}$. The perfect fidelity is indicated by a blue pentagram with the optimal time vector $\mathbf{t}_{\mathrm{Opt}}=[\tau_1,\tau_2,\tau_3]=[0.6505,0.2043,0.2978]$. By investigating the landscape with durations $T<\tau_{\mathrm{QSL}}$, perfect fidelity cannot be reached. It is interesting to notice that there is only one maximum, which is also the global maximum, in the landscape of fidelity. Therefore, all search points should converge towards the optimal time vector, and indeed this is true in our numerical optimization process. 100 search points are optimized using SD, and the maximal fidelity obtained after $10^6$ iterations is $F=1-2.220\times 10^{-16}$. The optimized time vector is $\mathbf{t}=[0.6504,0.2046,0.2976]$. Therefore we numerically verified $T_2^{\mathrm{min}}=\tau_{\mathrm{QSL}}$ with high accuracy.

	One of the benefits using SD optimizing over first class of bang-off controls is that we could estimate the number of local extremes and their values, as well as the location of corresponding controls. For instance, in Fig.~\ref{fig:PN0}(a) we show the quantum control landscape of log-Bures distance $\log_{10}(d_B)$ for type $P_{t_1}N_{t_2}0_{t_3}$ with $t_1+t_2+t_3=\tau_{\mathrm{QSL}}$. Note that perfect fidelity cannot be reached for type $P_{t_1}N_{t_2}0_{t_3}$ with total duration being $\tau_{\mathrm{QSL}}$. Two local minima exist in the landscape, and are estimated by optimizing 100 search points which are randomly sampled in the beginning of optimization. It is clearly shown in Fig.~\ref{fig:PN0}(b) two branches of search points saturate towards different values of $\log_{10}(d_B)$. It is worth noting that since the two local minima are far from each other, given that the step-size of optimization is small, it is hardly possible for search points to jump from one local trap to another. If the distance between two local minima is small, the search points might jump from one to another. Similar results are obtained for other types. 
	
	Therefore, we have estimated that $T_2^{\mathrm{min}}=\tau_{\mathrm{QSL}}$ and the perfect fidelity is reached for control fields $\{ P_{\tau_1}0_{\tau_2}N_{\tau_3},P_{\tau_1}0_{\tau_2}P_{\tau_3}\}$.
	
	\subsubsection{First class with $N_s=3$}
	
	\begin{figure}
		\includegraphics[width=1\linewidth, trim= 100 200 100 250,clip]{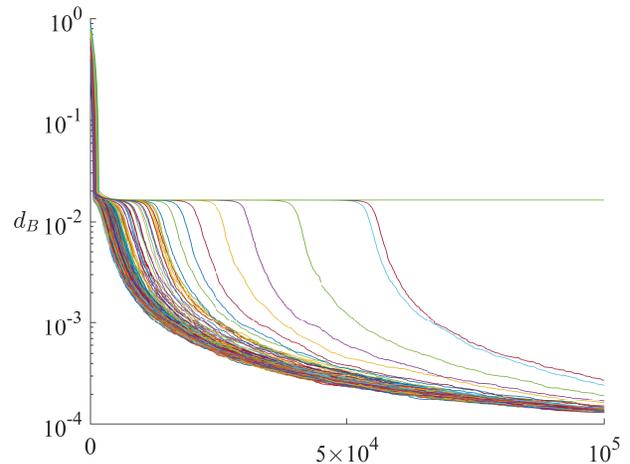}
		\caption{Bures distance $d_B$ versus iterations for type $P_{t_1}0_{t_2}N_{t_3}0_{t_4}$with the total evolution duration $T=\tau_{\mathrm{QSL}}$. 200 initial search points are optimized using SD with $10^5$ iterations.
		}
		\label{fig:P0N0}
	\end{figure}
	
	To further verify the time-optimal controls are of types with $N_s=2$, we continue to study the first class bang-off controls with $N_s=3$. For illustration, we consider the type $P_{t_1}0_{t_2}N_{t_3}0_{t_4}$. We would demonstrate that $T_3^{\mathrm{min}}=T_2^{\mathrm{min}}$, and one of the durations is zero $t_4=0$ .  
	
	By investigating the control landscape, we verify that the time vector corresponding the perfect fidelity with minimal duration is $\mathbf{t}_{\mathrm{Opt}}=[\tau_1,\tau_2,\tau_3,0]$ (data not shown). This is further confirmed by the optimization. Two hundred initial time vectors $t=[t_1,t_2,t_3,t_4]$ are optimized using SD with $t_1+t_2+t_3+t_4=\tau_{\mathrm{QSL}}$ for $10^5$ iterations. In Fig.~\ref{fig:P0N0} we show Bures distance $d_B$ as a function of iteration. It is interesting to notice from Fig.~\ref{fig:P0N0} that there are one local minimum and one global minimum of landscape, and they are not far from each other. Therefore, some of the search points are able to jump outside the local minimum and move towards the global one. 
	
	The best fidelity obtained using quasi-Newton is $F=1-\mathcal{O}(10^{-16})$, and the optimized time vector is $\mathbf{t}=[0.6505, 0.2043,0.2977, 4.064\times10^{-4}]$ with one duration being approximately equal to zero $t_4 \approx 0$. Similar results are obtained for type $\{P_{t_1}0_{t_2}N_{t_3}P_{t_4}, P_{t_1}0_{t_2}P_{t_3}0_{t_4}, P_{t_1}0_{t_2}P_{t_3}N_{t_4}\}$, while the perfect fidelity cannot be reached for other types with the total duration being $\tau_{\mathrm{QSL}}$. Therefore, we have numerically checked that $T_3^{\mathrm{min}}=T_2^{\mathrm{min}}=\tau_{\mathrm{QSL}}$ with high accuracy, hence the optimal controls with $N_s=3$ reduce to the ones with $N_s=2$.
	
	Considering that the optimal controls with $N_s=3$ reduce to the ones with $N_s=2$ and $T_3^{\mathrm{min}}=T_2^{\mathrm{min}}$, there is no need to study the first class of controls with larger switch numbers $N_s \geq 4$.
	
	In summary we have numerically verified for the transition considered in this section the time-optimal controls are $\{P_{\tau_1}0_{\tau_2}N_{\tau_3},P_{\tau_1}0_{\tau_2}P_{\tau_3}\}$, and the quantum speed limit is $\tau_{\mathrm{QSL}}$, with high precision.

	\subsubsection{Second class of bang-off control}
	
	\begin{figure}
		\includegraphics[width=1\linewidth, trim= 80 180 100 200,clip]{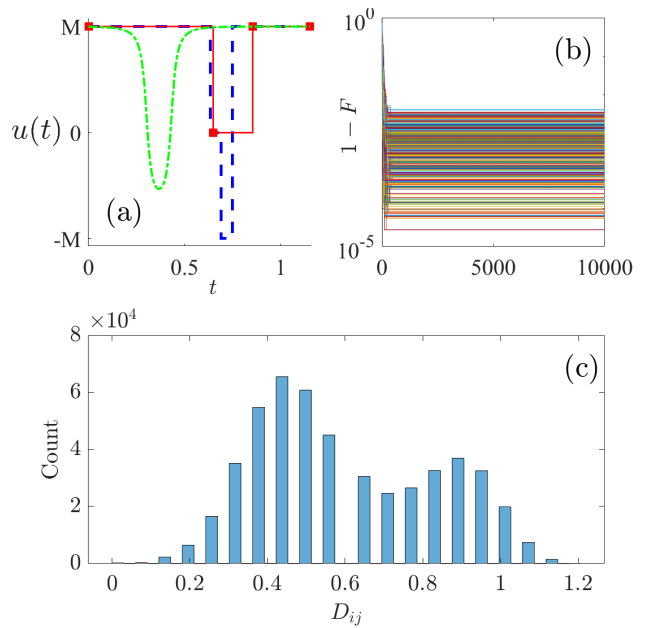}
		\caption{The results obtained for transition from $|0\rangle$ to $|\psi_E\rangle$ using 1-flip SD optimizing the second class of bang-off controls with total duration being $T=\tau_{\mathrm{QSL}}$. The number of time slots is fixed to be $N_T=20$. (a)The optimized control field found with first class of bang-off (red,solid) which is $P_{\tau_1}0_{\tau_2}P_{\tau_3}$, that with second class (blue, dashed), and with CRAB (grean, dash-dotted line). (b) Infidelity $1-F$ versus iterations with 1000 search points. (c) Distribution of distance $D_{ij}$ between optimized control fields. 
		}
		\label{fig:distance_2}
	\end{figure}

	Same as Sec.~\ref{sec:case1} we investigate the performance of the second class of bang-off controls using the 1-flip SD. 1000 initial control vectors are randomly generated and are optimized for $10^4$ iterations with the number of time slots being $N_T=20$. Distance between optimized control fields $D_{ij}$ are calculated.   
	
	In Fig.~\ref{fig:distance_2} the results of optimization are shown with total duration $T=\tau_{\mathrm{QSL}}$. In Fig.~\ref{fig:distance_2}(a) one of the time-optimal control fields $P_{\tau_1}0_{\tau_2}P_{\tau_3}$ is marked by the red solid line, and the optimized control field with  The temporal shape of the optimized control field found by 1-flip SD is again similar to that of time-optimal control $P_{\tau_1}0_{\tau_2}P_{\tau_3}$, except for certain duration the optimized control takes the value $u(t)=-M$, rather than the expected value $u(t)=0$. However, the best fidelity found by second class of bang-off is $F=1-2.77\times 10^{-6}$. In addition, we optimize the CRAB control field, the best fidelity obtained is $F=1-\mathcal{O}(10^{-13}) $ with $N_c=2$. The corresponding control field is indicated by green dash-dotted line,
	
	In Fig.~\ref{fig:distance_2}(b) we plot the infidelity $1-F$ as a function of optimization iteration. 1-flip SD stops updating after about 100 iterations. In Fig.~\ref{fig:distance_2}(c) distances between optimized control fields $D_{ij}$ form a bimodal distribution, which indicates the optimized control fields cluster around two time-optimal controls $\{P_{\tau_1}0_{\tau_2}P_{\tau_3},P_{\tau_1}0_{\tau_2}N_{\tau_3}\}$.
	
	Concerning the best fidelity obtained, the performance of first class of bang-off is much better than CRAB and second class, while second class of bang-off performs worst.

	\section{Three-level system}\label{sec:threelevel}
	\begin{figure}
		\includegraphics[width=0.8\linewidth, trim= 100 220 100 250,clip]{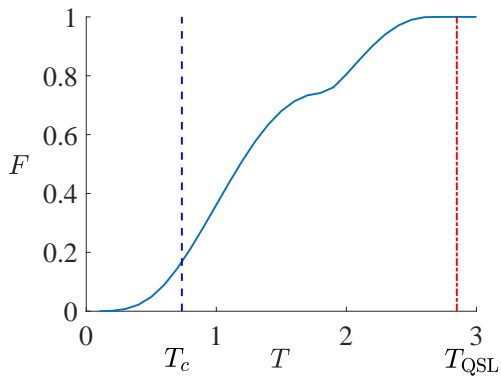}
		\caption{Fidelity $F$ versus total duration $T$ in three-level system obtained with the first class of bang-off control field. Two critical time points are $T_c$ and $T_{\mathrm{QSL}}$. For $T\leq T_c$, the optimal control is $P_T$ (and $N_T$); for $T > T_c$ the optimal control is of type $N_s=4$.
		}
		\label{fig:fidelity_3level}
	\end{figure}	
	
	To test the performance of the systematic scheme proposed, we apply it in the three-level quantum system. Specifically, we consider the Hamiltonian of a three-level quantum system described in the following:
	\begin{equation}
		H(h(t))  =-E H_0 + \mu_1 H_1 + \mu_2 u(t) H_2,
		\label{eq:threelevel}
	\end{equation}
	where the drift Hamiltonian $H_0$ and $H_1$, and the control Hamiltonian $H_2$ are
	\begin{equation}	
		\centering {
			\begin{matrix}
				H_0=\begin{bmatrix}
					1 & 0 & 0\\0 & 0 & 0\\0 & 0 & -1
				\end{bmatrix} ,
				H_1=\begin{bmatrix}
					0 & 1 & 0\\1 & 0 & 0\\0 & 0 & 0
				\end{bmatrix} , 
				H_2 =\begin{bmatrix}
					0 & 0 & 0\\0 & 0 & 1\\0 & 1 & 0
				\end{bmatrix},
			\end{matrix}
		}
	\end{equation}
	respectively. $H_1$ couples state $|1 \rangle $ and $|2 \rangle $, and  $H_2$ couples state $|2 \rangle $ and $|3 \rangle$, where $|1 \rangle = [1,0,0]^{\mathrm{T}}$, etc. 
	
	Without loss of generality, we set $E_0=1$, the coupling strengths $\mu_1=1$, and $\mu_2=2$. The control field is constrained, i.e., $-1\leq u(t) \leq 1$. We wish to drive the system from the initial state $|1 \rangle $ to the target state $|3 \rangle$.
	
	It is worthy noting that 
	
	\begin{equation}
		\langle 3| H^n(h) |1 \rangle =-\langle 3| H^n(-h) |1 \rangle, 
		\label{eq:mirrorsymmetry}
	\end{equation}
	where $n$ is integer. Therefore, it is straightforward to show that there is a mirror symmetry $F(h(t),T)= F(-h(t),T)$:
	
	\begin{eqnarray}
		F(h(t),T)&=&|\langle 3|\; \mathcal{T}\exp (- \mathrm{i}\int_0^T H[h(t)]  \mathrm{d}t \;|1\rangle|^2 \nonumber\\
		&=&|\langle 3| \; \mathcal{T}\exp (- \mathrm{i}\int_0^T H[-h(t)]  \mathrm{d}t\; |1\rangle|^2 \nonumber\\
		&=& F(-h(t),T).
		\label{eq:symmetry}
	\end{eqnarray}
	This symmetry can help reduce by half the number of types to be calculated.
	
	In Fig.~\ref{fig:fidelity_3level} we show the best fidelity obtained as a function of the total duration $T$ by optimizing the first class of bang-off controls. Two critical time points are shown, i.e., $T_c$ and $T_{\mathrm{QSL}}$. For different total duration $T$, the corresponding optimal control is in general different. For $T\in [0, T_c]$ the optimal control is of type $N_s=0$, and $T\in (T_c, T_{\mathrm{QSL}}]$ the optimal control is of type $N_s=4$.
	
	\subsection{Optimal control for $T\in [0, T_c]$}
\begin{figure}
		\includegraphics[width=1\linewidth, trim= 100 220 100 240,clip]{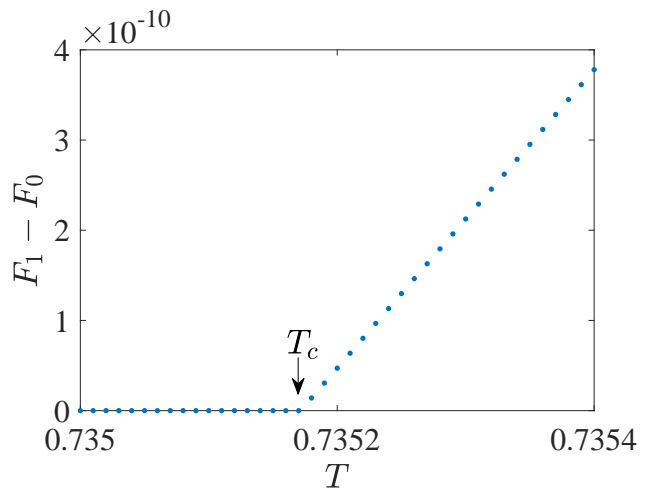}
		\caption{$F_1-F_0$ versus total duration $T$. $F_1$ is the best fidelity obtained with $N_s=1$, and $F_0$ with $N_s=0$. $F_1-F_0 =0$ when $T \leq T_c$, and $F_1-F_0 >0$ when $T > T_c$. $T_c=0.73517$ is indicated by an arrow. 
		}
		\label{fig:TC}
	\end{figure}

	To determine the optimal control for given $T$, we optimize over all types of the first class of bang-off control. It is found that for $T\leq 0.73157$, the best fidelity obtained with $N_s\geq 1$ is equal to that obtained with $N_s=0$, while for $T> 0.73157$ the former is larger than the latter; cf. Fig.~\ref{fig:TC}. For $N_s=0$ the best fidelity is obtained with both $P_T$ and $N_T$, which is implied by Eq.~(\ref{eq:mirrorsymmetry}). In addition, for $T\leq 0.73157$ the optimal control fields obtained with $N_s\geq 1$ are either $P_T$ or $N_T$. 
	
	\begin{figure}
		\includegraphics[width=1\linewidth, trim= 100 220 100 250,clip]{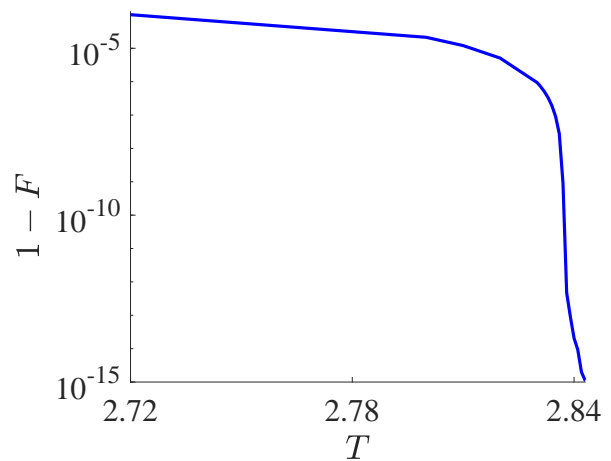}
		\caption{Infidelity $1-F$ versus total duration $T$ obtained with $N_s=4$ bang-off control, which is the estimation 	of optimal control for $T \geq T_c$. The $y$-axis is in $\log$-scale.  
		}
		\label{fig:infidelity}
	\end{figure}
	
	In Ref.~\cite{Bukov2018b} the region with $T\in [0, T_c]$ is called \textit{overconstrained} since the landscape in this region is convex such that the search for local minimum is easy. By observing the behavior of correlator and other functions, the value of $T_c$ is roughly estimated~\cite{Bukov2018b}. By using our method, it is much easier to determine the value of $T_c$ with very high accuracy. For the three-level quantum system considered here, we compare the best fidelity obtained with $N_s=1$ and that with $N_s=0$, the value of $T_c$ is located by checking the duration $T$ with which $F_1-F_0$ becomes nonzero. See Fig.~\ref{fig:TC} for more information.  
	
	\subsection{Optimal control for $T\in (T_c, T_{\mathrm{QSL}}]$}

	As $T$ increases and is larger than $T_c$, the optimal control is not $P_T$ or $N_T$ anymore. For $T>T_c$, we optimize over the first class of bang-off control with different value of $N_s$. It is found that the best fidelity obtained with $N_s=4$ is larger than those obtained with $N_s\leq 3$, and is equal to those obtained with $N_s \geq 5$. In addition, the optimal control fields obtained with $N_s \geq 5$ are reduced to the ones with $N_s=4$.
	\begin{figure}
		\includegraphics[width=1\linewidth, trim= 100 220 100 260,clip]{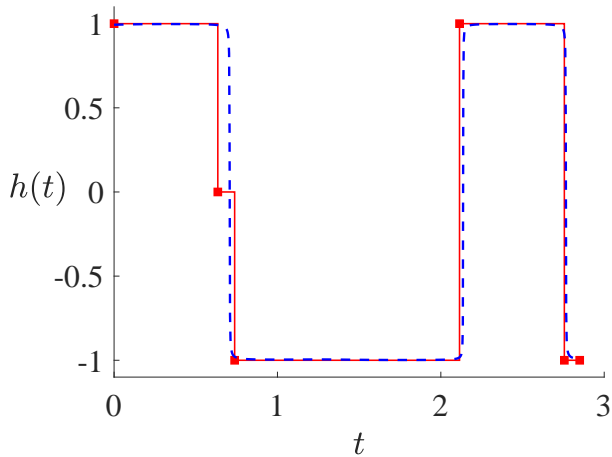}
		\caption{One of the optimal control fields $h(t)$ obtained by the first class of bang-off control (red, solid) 	$P_{t_1}0_{t_2}N_{t_3}P_{t_4}N_{t_5}$ and by CRAB (blue, dashed), with the total duration being the estimated 		quantum speed limit $T=2.850 \approx T_{\mathrm{QSL}} $.
		}
		\label{fig:control}
	\end{figure}	

    We conjecture that for $T\in (T_c, T_{\mathrm{QSL}}]$ the optimal control field is bang-off control with $N_s=4$. The reasons are as follows. In Ref.~\cite{Bukov2018b} for total duration in the overconstrained region and glassy region, the optimal control is of bang-off type, which is obtained by averaging one thousand optimized bang-bang controls. This is further verified in Ref.~\cite{Li2022} by employing the systematic method proposed in this paper. The number of switches of the bang-off control is small, i.e., $N_s=1$ ($N_s=2$) for overconstrained (glassy) region. However, for total duration in the symmetry-broken region, the optimal control field obtained is continuous, rather than bang-off control which is piece-wise constant~\cite{Bukov2018b}. This is also reflected in Ref.~\cite{Li2022} through the fact that the number of switch of the optimized bang-off control in this region is $N_s=9$. The bang-off control with such a large number of switch serves as a reasonable approximation to the continuous control. In such case, $u(t)$ takes values between the maximal and minimal value, and is in general referred to as 'singular'~\cite{Boscain2006,Brady2021}. Note that in the case of two-level quantum system considered here, the singular control is in fact off control~\cite{Boscain2006}. In general, however, this conclusion does not hold in multiple-level quantum system~\cite{Brady2021}. Given the fact that for $T\in (T_c, T_{\mathrm{QSL}}]$ the optimized bang-off control is of type with $N_s=4$, which is small, and the fact that the optimized control field obtained with CRAB is very similar to the bang-off control (see Fig.~\ref{fig:control}), we conjecture that the optimal control in this region is bang-off control with $N_s=4$.

	From the numerical results we observed that there are eight types of optimal control fields in this region, i.e., $P0NPN$, $PNPNP$, $PN0PN$, $0NPNP$, and other four types are the negative of the former four types, see Eq.~(\ref{eq:symmetry}). In Fig.~\ref{fig:infidelity} we show the infidelity $1-F$, obtained with $N_s=4$ optimal control field, as a function of the total duration. A vertical asymptote of infidelity is present around $T\approx 2.84$. Furthermore, the unit fidelity $F=1-\mathcal{O}(10^{-16})$ is obtained with $T=2.850$. Therefore, we estimate the the quantum speed limit to be $T_{\mathrm{QSL}} \approx 2.850$. 
	
	In Fig.~\ref{fig:control} we show one of the time-optimal controls obtained numerically with type $P_{t_1}0_{t_2}N_{t_3}P_{t_4}N_{t_5}$. As a comparison, we employ the CRAB method to optimize the control field for $T=2.850$ with various values of $N_c$. The best fidelity obtained is $F=1-3.996\times 10^{-6}$ with $N_c=3$. It is worth noting that the optimized control field obtained with CRAB approximates the bang-bang control, and is very similar to the one obtained with bang-off control. However, the infidelity obtained with CRAB is \textit{ten} orders of magnitude greater than the one obtained with bang-off control. Therefore, the performance of our method of bang-off is much better than that of CRAB.
	
	\section{Conclusions}\label{sec:conclusions}
	
	We have proposed a systematic scheme for numerically estimating the quantum speed limit and approximating the temporal shape of optimal control with bounded amplitude. It has been studied as examples the optimal control problems of quantum state transition processes in two-level and three-level quantum system. 
	
	For the two-level quantum system, the numerical results are compared to the analytic solutions. We found that the deviations from analytic solutions are significantly small. We have also investigated the robustness of first class of bang-off controls. In addition, the performance of first class is better than that of second class and that of CRAB, concerning the fidelity obtained and the optimized control field found. 
	
	For the three-level quantum system, we study the control problem with real control field with bounded amplitude. Optimal controls for different regions of total duration are obtained. QSL and time-optimal control are estimated. We found that although the optimized control field obtained using CRAB is very similar to the one obtained using our method, the infidelity obtained in the former is much greater than the latter with the total duration being QSL. 
	
	Since the temporal shape of the first class of bang-off control is simple, it is easy to implement such control fields in experiment. Our scheme might shed light on quantum optimal control problems in which the analytical solution is absent.
	

	%

\end{document}